\documentclass[a4paper,12pt]{article}

\newcommand{\sect}[1]{\setcounter{equation}{0}\section{#1}}

\newcommand{\bea}{\begin{eqnarray}}
\newcommand{\eea}{\end{eqnarray}}
\newcommand{\be}{\begin{equation}}
\newcommand{\ee}{\end{equation}}
\newcommand{\vs}[1]{\vspace{#1 mm}}

\newcommand{\dsl}{\pa \kern-0.5em /}

\newcommand{\pa}{\partial}

\begin{document}
\topmargin 0pt
\oddsidemargin 0mm

\begin{flushright}
\end{flushright}

\vs{2}
\begin{center}
{\Large \bf  
Penrose limit and NCYM theories in diverse dimensions}
\vs{10}

{\large Somdatta Bhattacharya and S. Roy}
\vspace{5mm}

{\em 
 Saha Institute of Nuclear Physics,
 1/AF Bidhannagar, Calcutta-700 064, India\\
E-Mails: som, roy@theory.saha.ernet.in\\}
\end{center}

\vs{5}
\centerline{{\bf{Abstract}}}
\vs{5}
\begin{small}
We obtain the Penrose limit of NCYM theories in dimensions $3 \leq d \leq 6$
which originate from (D$(p-2)$, D$p$) supergravity bound state
configurations for $2 \leq p \leq 5$ in the so-called NCYM limit. In most
cases the Penrose limit does not lead to solvable string theories except for
six-dimensional NCYM theory. We obtain the masses of various bosonic 
coordinates and observe that they are light-cone time dependent and their
squares can be negative as has also been observed in other cases in 
the literature.
When the non-commutative effect is turned off we recover the results of
Penrose limit of ordinary D$p$-branes in the usual YM limit. We point out
that for $d = 6$ NCYM theory, there exists another null geodesic in the
neighborhood of which the Penrose limit leads to a solvable 
string theory. We briefly discuss the quantization of this theory and show
that this pp-wave background is half supersymmetric.
\end{small}

\sect{Introduction}

String theory in NSNS and RR pp-wave backgrounds has generated a lot of 
interest recently mainly because it has maximal supersymmetry for type
IIB theory \cite{bfhp} (like flat Minkowski space and AdS$_5 \times$S$^5$), 
the
corresponding GS action in the light-cone gauge is exactly solvable 
\cite{met,metsey,rusey} and
can be obtained by taking a Penrose limit \cite{pen}\footnote{The idea that any
space-time obtained as solution of Einstein's equation can be reduced
by Penrose limit to pp-wave background has been extended by G\"uven 
\cite{guv} for supergravity field equation including a $p$-form
field strength.} on a suitably 
chosen coordinates
in the neighborhood of an appropriate null geodesic of AdS$_5 \times$S$^5$
\cite{blfhp}. 
Since the Maldacena conjecture \cite{mal,agmoo,gkp,witt} relates type 
IIB string theory on AdS$_5
\times$S$^5$ to the large $N$, ${\cal N} = 4$, $d = 4$, $SU(N)$ gauge 
theory, taking a Penrose limit amounts to going to a particular subsector
of the gauge theory \cite{bmn}. In this subsector even though the 
't Hooft coupling
$g_{YM}^2 N$ is large as $N \to \infty$, there is an effective 't Hooft
coupling $g_{YM}^2N/J^2$ which remains fixed. Here $J$ is the R-charge of
a $U(1)$ subgroup of the full $SO(6)$ R-symmetry group of the ${\cal N} 
= 4$ gauge theory and scales as $J \sim \sqrt{N}$. As string theory in this
case is exactly solvable, one uses this information to compute the
anomalous dimensions of a particular set of operators in that subsector of
gauge theory, corresponding to massive string modes. This is not possible
for the AdS$_5 \times$S$^5$ background before taking the Penrose limit simply
because the 't Hooft coupling is divergent. This has led BMN \cite{bmn} 
to conjecture
an exact correspondence between string theory in pp-wave background (beyond
supergravity) and the subsector of gauge theory mentioned above and
has been generalized to many other AdS/CFT-like examples.

While the near horizon geometry of D3-branes \cite{mal,agmoo} 
(which is AdS$_5 \times$S$^5$)
is related to conformal field theory on the D3-brane world volume, the near
horizon limits of other D$p$-branes \cite{imsy} are related to 
non-conformal theories on
the world volume. It is of interest to see whether Penrose limit for the
latter class of theories \cite{bfp,ps} can teach us something new. 
In ref.\cite{gps}, it has
been pointed out that Penrose limit on the near horizon geometry of 
D$p$-branes leads to time dependent pp-wave background of string theory.
The string theories in these cases are not exactly solvable. However, for 
these classes of systems there is an intriguing connection between the
associated time-dependent quantum mechanical problem and the RG flow in the
dual gauge theory \cite{gps,chkw,bjlm}\footnote{See also \cite{fis} for some
recent discussion on the possibility of obtaining solutions to a large class
of string theories in pp-wave background with time dependent masses.}. 
The Penrose limit of the 
near horizon geometry 
of NS5-branes,
the supergravity dual of LST, is the Nappi-Witten background \cite{nw}
and gives 
an exactly
solvable string theory. This has been studied in \cite{go,kirp,hrv}. Another 
class of 
non-local theories, namely, NCYM in (3+1) dimensions which originate from
(D1, D3) bound state in the NCYM limit \cite{hashi,mr} and its Penrose 
limit has also been
studied in ref.\cite{hrv}. However, four dimensional NCYM theory 
does not lead to
a solvable string theory. Recently, the Penrose limit of OD5 theory 
\cite{gmss} originating
from (NS5, D5) bound state \cite{luroy} in the so-called OD5 limit 
\cite{aor,mitro} has been studied.
Interestingly, it gives a solvable string theory. The string spectrum and 
their relation to the states in the six-dimensional ``gauge'' theory has
been discussed in \cite{osa}. The Penrose limit of all OD$p$ theories, 
four dimensional
NCOS and OM theory has been discussed in \cite{aliku}.

In this paper we study the Penrose limit of NCYM theories \cite{aos,lur}
(actually, their 
supergravity duals to be precise) in dimensions $3 \leq d \leq 6$. These
theories in various dimensions arise from (D$(p-2)$, D$p$) bound state
(with $2 \leq p \leq 5$) of type II supergravities in the so-called NCYM 
limit. We obtain the Penrose limits for these supergravity configurations
by making suitable coordinate changes in the neighborhoods of  null geodesics
following ref.\cite{hrv}. A suitable scaling parameter in terms of number of 
D$p$-branes, coupling constant and the noncommutativity parameter of the NCYM
theories is chosen to obtain the Penrose limit. This gives a one parameter
($l$) family of string theories in a time dependent 
pp-wave background. However,
because of the time dependence they are not solvable. For $p=3$, we get
the Penrose limit of four dimensional NCYM theory obtained in ref.\cite{hrv},
which reduces to the maximally supersymmetric Penrose limit of AdS$_5
\times$S$^5$ when the noncommutative effect is turned off. For general $p$
we compute the mass$^2$'s of the various bosonic fields and find that they
can be negative (in both IR and UV) in some cases signalling a possible quantum
mechanical instability for the corresponding string theory. When the  
noncommutativity parameter approaches zero (or in the IR) we get back
the Penrose limit of ordinary D$p$-branes obtained in ref.\cite{gps}. 
By examining
the evolution equation along the geodesic we find that in six dimensions
the Penrose limit of NCYM theory can give a solvable string theory when the 
parameter `$l$' is saturated i.e. $l=1$. In this case there exists a null 
geodesic very similar to the AdS$_5 \times$S$^5$ case and the Penrose limit 
gives
a string theory where two of the eight bosonic coordinates have 
non-zero time independent masses. We briefly discuss the quantization of this
theory and show that the background
is half supersymmetric.

This paper is organized as follows. In section 2, we briefly review the
(D$(p-2)$, D$p$) supergravity bound state in the NCYM limit. The Penrose
limit of the NCYM theories in various dimensions is discussed in section 3.
In section 4, the (D5, D3) system and the Penrose limit of 6D NCYM
theory is discussed separately which leads to a solvable string theory. Our
conclusion is presented in section 5.

\sect{(D$(p-2)$, D$p$) bound state and the NCYM limit}

In this section we give a brief review of (D$(p-2)$, D$p$) supergravity
bound state configuration for $2 \leq p \leq 5$ and their NCYM 
limit\footnote{For details see for example ref.\cite{lur}.}. These are the
supergravity duals of NCYM theories in dimensions $3\leq d \leq 6$. The 
string metric, the dilaton and the NSNS two-form potential have the 
forms\footnote{We do not give the RR-forms, as they are not important
for our purpose here and can be found, for example, in \cite{bmm}. See also
\cite{cp}.},
\bea
ds^2 &=& H^{1/2}\left[H^{-1}(-(d\tilde{x}^0)^2 + \sum_{i=1}^{p-2}
(d\tilde{x}^i)^2) + H'^{-1}((d\tilde{x}^{p-1})^2 + (d\tilde{x}^p)^2
) + d\tilde{r}^2 + \tilde{r}^2 d\Omega_{8-p}^2\right]\nonumber\\
e^{\phi} &=& g_s \frac{H^{(5-p)/4}}{H'^{1/2}}\nonumber\\
B &=& \tan\varphi H'^{-1} d\tilde{x}^{p-1} \wedge d\tilde{x}^p
\eea
Here $H$ and $H'$ are two harmonic functions defined as
\be
H = 1 + \frac{Q_p}{\tilde{r}^{7-p}}, \qquad H' = 1 + \frac{\cos^2\varphi Q_p}
{\tilde{r}^{7-p}}
\ee
with $Q_p = g_s c_p \sqrt{n^2 + m^2} \alpha'^{(7-p)/2}$, $c_p = 2^{5-p}
\pi^{(5-p)/2} \Gamma(\frac{7-p}{2})$. Also $g_s$ is the string coupling 
constant. $n$ is the number of D$p$-branes and $m$ is the number of 
D$(p-2)$-branes per $(2\pi)^2 \alpha'$ of two codimensional area of 
D$p$-branes. $\tilde{r}$ is the radial coordinate transverse to D$p$-branes
and $d\Omega_{8-p}^2$ is the line element of unit $(8-p)$ dimensional sphere.
The angle $\varphi$ is defined as,
\be
\cos\varphi = \frac{n}{\sqrt{n^2 + m^2}}
\ee
The charge $Q_p$ above can therefore be written as $Q_p = (g_s c_p n 
\alpha'^{(7-p)/2})/\cos\varphi$ in terms of which the harmonic functions
in (2.2) take the following forms,
\be
H = 1 + \frac{ng_s c_p \alpha'^{(7-p)/2}}{\cos\varphi \tilde{r}^{7-p}},
\qquad H' = 1 + \frac{n g_s c_p \alpha'^{(7-p)/2}\cos\varphi}
{\tilde{r}^{7-p}}
\ee
The NCYM limit is defined by scaling $\alpha' \to 0$, keeping the following
quantities fixed,
\be
r = \frac{\tilde{r}}{\alpha'}, \qquad b = \frac{\alpha'}{\cos\varphi},
\qquad g = g_s \alpha'^{(p-5)/2}
\ee
where $r$ is the energy parameter and $b$ is the noncommutativity parameter
in the NCYM theory. The NCYM gauge coupling $\bar{g}_{YM}^2$ is related
to $g$ and $b$ as $\bar{g}_{YM}^2 = (2\pi)^{p-2} g b$. Substituting (2.5)
in (2.1) we get the NCYM supergravity configuration as follows,
\bea
ds^2 &=& \alpha'\frac{(ar)^{\frac{7-p}{2}}}{b}\left[-(dx^0)^2 + 
\sum_{i=1}^{p-2}
(dx^i)^2 + \frac{1}{1+(ar)^{7-p}}((dx^{p-1})^2 + (dx^p)^2
) \right.\nonumber\\
& &\qquad\qquad\qquad\qquad\qquad\qquad\qquad\qquad\left.+ 
\frac{b^2}{(ar)^{7-p}}
(dr^2 + r^2 d\Omega_{8-p}^2)\right]\nonumber\\
e^{\phi} &=& g b^{\frac{5-p}{2}} \frac{(ar)^{\frac{(7-p)(p-3)}{4}}}
{\sqrt{1 + (ar)^{7-p}}}\nonumber\\
B &=& \frac{\alpha'}{b}\frac{(ar)^{7-p}}{1+(ar)^{7-p}} dx^{p-1} 
\wedge dx^p
\eea
In the above we have defined the fixed coordinates as,
\bea
x^{0,1,\ldots,p-2} &=& \tilde{x}^{0,1,\ldots,p-2}\nonumber\\
x^{p-1,p} &=& \frac{b}{\alpha'}\tilde{x}^{p-1,p}
\eea
In terms of the fixed coordinates the noncommutativity parameter $b$ is given
by $[x^{p-1},\, x^p] = i b$ and $a$ is defined as $a^{7-p} = b/(nc_pg)$. The 
NCYM effective gauge coupling is $g_{\rm eff}^2 \sim \bar{g}_{YM}^2 n r^{p-3}$.
The supergravity description (2.6) is valid when the curvature in units of
$\alpha'$ i.e. $\alpha'{\cal R} \sim g_{\rm eff}^{-1} \ll 1$ and the dilaton
$e^{\phi} \ll 1$. The curvature condition implies
\be
g_{\rm eff}^2 \gg 1 \quad \Rightarrow \quad \cases {r \ll (gbn)^{\frac{1}
{3-p}} &  for $\quad p<3$\cr
r \gg (\frac{1}{gbn})^{\frac{1}{p-3}} & for $\quad p>3$}
\ee
and the dilaton condition implies
\be
e^{\phi} \ll 1 \quad \Rightarrow \quad g^2 b^{5-p} 
(ar)^{\frac{(7-p)(p-3)}{2}} \ll
1+(ar)^{7-p}
\ee
In the UV, $ar \gg 1$ and this implies $e^\phi \sim (ar)^{(7-p)(p-5)/4}$
which vanishes for $p < 5$ and for $p = 5$, $e^\phi \sim g$ which will remain
small if $g \ll 1$. In the IR when $ar \ll 1$, we note that the supergravity 
configuration reduces precisely to D$p$-branes in the near horizon limit
i.e. ordinary YM theory if we identfy,
\be
\frac{a^{\frac{7-p}{2}}}{b} = \frac{1}{\sqrt{g_{YM}^2 c_p n}} = \frac{1}{
\sqrt{gbc_pn}}
\ee
where $g_{YM}^2$ is now the gauge coupling of ordinary YM theory. Note
from (2.8) that for $p<3$, NCYM theory is UV free, whereas for $p>3$, the
field theory breaks down and we need new degrees of freedom.

\sect{Penrose limit and NCYM theories}

In order to take Penrose limit of the NCYM supergravity configuration (2.6),
we first rescale the coordinates $x^{0,1,\ldots,p} \to (b/a) 
x^{0,1,\ldots,p}$. Then defining the scaling parameter
\be
R^2 = \alpha' \frac{b}{a^2}
\ee
we rewrite the metric in (2.6) as,
\bea
ds^2 &=& R^2 e^{\frac{7-p}{2}U}\left[-(dx^0)^2 + 
\sum_{i=1}^{p-2}
(dx^i)^2 + \frac{1}{1+e^{(7-p)U}}((dx^{p-1})^2 + (dx^p)^2
) \right.\nonumber\\
& &\qquad\qquad\qquad\left.+ 
e^{-(5-p)U}
(dU^2 + \cos^2\theta d\psi^2 + d\theta^2 +\sin^2\theta d\Omega_{6-p}^2)\right]
\eea
Here we have defined a new variable $e^U = (ar)$ and have written 
$d\Omega_{8-p}^2 = \cos^2\theta d\psi^2 + d\theta^2 + \sin^2\theta 
d\Omega_{6-p}^2$. Now we consider a null geodesic in the $(x^0,\,U,\,\psi)$
plane and so, we have for the geodesic $x^1 = x^2 = \cdots = x^p = 0$,
$\theta = 0$. The effective Lagrangian associated with this geodesic has the 
form,
\be
{\cal L} = - e^{\frac{7-p}{2}U} ((x^0)')^2 + e^{\frac{p-3}{2}U} (U')^2 +
e^{\frac{p-3}{2}U} (\psi')^2
\ee
where `prime' denotes the derivative with respect to the affine parameter
`$u$' along the geodesic. Since the Lagrangian does not depend on $x^0$
and $\psi$, we get the following constants of motion,
\be
e^{\frac{7-p}{2}U} (x^0)' = E, \qquad e^{\frac{p-3}{2}U} \psi' = J
\ee
For null geodesic, substituting (3.4) to (3.3) and equating it to zero,
we get the evolution equation for $U$ as,
\be
U' = \frac{\sqrt{1-e^{(5-p)U}l^2}}{e^U}
\ee
where we have defined $J/E = l$ and also have scaled the affine parameter
by $E$. From (3.5) we find that $l^2 \leq e^{(p-5)U}$. We remark that
for $p=5$ and $l=1$, $e^U\, =$ constant is a solution to the evolution 
equation. We will use this while discussing the Penrose limit 
and six dimensional
NCYM theory. For future reference we note that the null geodesic in this case
can be restricted to $(x^0,\,\psi)$ plane. However, for $p\neq 5$, it is
clear from (3.5) that such a null geodesic does not exist for any value
of $l$. In other words, the null geodesic can not stay in the $(x^0,\,\psi)$
plane even if we take $l=1$ or 0. The parameter $l$ has a clear geometric
meaning, namely, it is the ratio of the angular momentum and the energy
of a test particle moving along the trajectory described by the effective 
Lagrangian (3.3). So, $l=1$ means that for such a particle restricted to the
trajectory the difference of energy and angular momentum must vanish. In
the Penrose limit, when both the energy and the angular momentum are scaled
as $R^2 \to \infty$, their difference remains finite and small in the
neighborhood of the null geodesic. This fact gets translated to the gauge 
theory side as the difference of the energy and the $U(1)$ R-charge of the
states being finite and small at a fixed energy scale (since $U$ is constant).
An analogous situation also occurs for the maximally supersymmetric AdS$_5
\times$S$^5$ case, where the string theory results give information about the
near BPS states in gauge theory. 

Eq.(3.5) can be integrated to obtain $u$ as,
\be
u = \int \frac{e^U}{\sqrt{1-e^{(5-p)U}l^2}} dU
\ee
This relation can then be inverted to express $e^U$ as a function of $u$ and 
let us call that function as $e^U = f(u)$. Now we make the following change of
variables from $(x^0,\,U,\,\psi) \to (u,\,v,\,x)$:
\bea
dU &=& \frac{\sqrt{1-f^{5-p}l^2}}{f} du\nonumber\\
dx^0 &=& \frac{1}{f^{(7-p)/2}} du + 2 dv + l dx\nonumber\\
d\psi &=& \frac{l}{f^{(p-3)/2}} du + dx
\eea
Substituting (3.7) into (3.2) and rescaling the coordinates as, $u \to u$,
$v \to v/R^2$, $\theta \to z/R$, $x \to x/R$, 
$x^{1,\ldots,p} \to x^{1,\ldots,p}/R$ and keeping others fixed along 
with $R \to \infty$, we
find
\bea
ds^2 &=& -4 du dv - f^{\frac{3-p}{2}} l^2 \vec{z}^2 du^2 + f^{\frac{p-3}{2}}
(1-l^2 f^{5-p}) dx^2 + f^{\frac{7-p}{2}}\sum_{i=1}^{p-2} (dx^i)^2 \nonumber\\
& & \qquad +\frac{f^{\frac{7-p}{2}}}{1+f^{7-p}}\sum_{j=p-1}^p (dx^j)^2
+ f^{\frac{p-3}{2}}(dz^2 + z^2 d\Omega_{6-p}^2)
\eea
where $\vec{z}^2 = z^2 = z_1^2 + \cdots + z_{7-p}^2$. This is the form of
the metric in Rosen coordinates. It can be written in Brinkman form if we
make another change of variables,
\bea
u & \rightarrow& x^+\nonumber\\
x^{1,\ldots,p-2} & \rightarrow& f^{\frac{p-7}{4}}x^{1,\ldots,p-2}\nonumber\\
x^{p-1,p} & \rightarrow& f^{\frac{p-7}{4}} \sqrt{1+f^{7-p}}x^{p-1,p}\nonumber\\
x & \rightarrow& \frac{f^{\frac{3-p}{4}}}{\sqrt{1-l^2 f^{5-p}}}x\nonumber\\
\vec{z} & \rightarrow& f^{\frac{3-p}{4}} \vec{z}\nonumber\\
v &\rightarrow& x^- - \frac{1}{8}\left[G_z \vec{z}^2 + G_{p-1,p}\sum_{j=p-1}^p
(x^j)^2 + G_{1,\ldots,p-2}\sum_{i=1}^{p-2}(x^i)^2 + G_x x^2\right]
\eea
where
\bea
G_z &=& \frac{(f^{\frac{p-3}{2}})'}{f^{\frac{p-3}{2}}}, \qquad
G_{p-1,p}\,\, =\,\, \frac{\left(\frac{f^{\frac{7-p}{2}}}{1+f^{7-p}}\right)'}
{\left(\frac{f^{\frac{7-p}{2}}}{1+f^{7-p}}\right)}\nonumber\\
G_{1,\ldots,p-2} &=& \frac{(f^{\frac{7-p}{2}})'}{f^{\frac{7-p}{2}}}, \qquad
G_x \,\,=\,\, \frac{[f^{\frac{p-3}{2}}(1-l^2 f^{5-p})]'}
{[f^{\frac{p-3}{2}}(1-l^2 f^{5-p})]}
\eea
Using (3.9), the metric in (3.8) takes the form,
\bea
ds^2 &=& -4 dx^+ dx^- + \left[F_z \vec{z}^2 + F_{p-1,p}\sum_{j=p-1}^p (x^j)^2
+ F_{1,\ldots,p-2}\sum_{i=1}^{p-2} (x^i)^2 + F_x x^2\right] 
(dx^+)^2\nonumber\\
& & \qquad\qquad + dx^2 + \sum_{k=1}^p (dx^k)^2 + d\vec{z}^2
\eea
where
\bea
F_z &=& -f^{3-p} l^2 + \frac{(f^{\frac{p-3}{4}})''}
{f^{\frac{p-3}{4}}}, \qquad
F_{p-1,p}\,\, =\,\, \frac{\left(\frac{f^{\frac{7-p}{4}}}
{\sqrt{1+f^{7-p}}}\right)''}
{\left(\frac{f^{\frac{7-p}{4}}}{\sqrt{1+f^{7-p}}}\right)}\nonumber\\
F_{1,\ldots,p-2} &=& \frac{(f^{\frac{7-p}{4}})''}{f^{\frac{7-p}{4}}}, \qquad
F_x \,\,=\,\, \frac{[f^{\frac{p-3}{4}}\sqrt{1-l^2 f^{5-p}}]''}
{[f^{\frac{p-3}{4}}\sqrt{1-l^2 f^{5-p}}]}
\eea
Here `prime' denotes derivative with respect to $u = x^+$. Eq.(3.11) is 
the NCYM supergravity metric in the Penrose limit in Brinkman form. The forms
of the dilaton and the NSNS two form in these coordinate system is given as,
\bea
e^{\phi} &=& g b^{\frac{5-p}{2}}\frac{f^{\frac{(7-p)(p-3)}{4}}}
{\sqrt{1+f^{7-p}}}\nonumber\\
B &=& f^{\frac{7-p}{2}} dx^{p-1} \wedge dx^p
\eea
The mass$^2$'s of various bosonic coordinates are given as $m_z^2 = - F_z$,
$m_{p-1,p}^2 = - F_{p-1,p}$, $m_{1,\ldots,p-2}^2 = -F_{1,\ldots,p-2}$ and
$m_x^2 = - F_x$ and they can be calculated from (3.12) as,
\bea
m_z^2 &=& \frac{7-p}{16} f^{3-p} \left[-(3-p) f^{p-5} + (p+1) 
l^2\right]\nonumber\\
m_{1,\ldots,p-2}^2 &=& \frac{7-p}{16} f^{3-p} \left[-(3-p) f^{p-5} + (13-3p) 
l^2\right]\nonumber\\
m_{p-1,p}^2 &=& \frac{(7-p) f^{3-p}}{16 (1+f^{7-p})^2} \left[-(3-p) f^{p-5}
(1+f^{7-p})^2 + (13-3p) l^2 (1+ f^{7-p})^2 \right.\nonumber\\
& &  + 4 (19-3p) f^2 (1 - f^{5-p}l^2)
(1+f^{7-p})
- 4(5-p) f^{7-p} l^2 (1+f^{7-p}) \nonumber\\
& & \left.- 12 (7-p) f^{9-p} 
(1-f^{5-p}l^2)\right]\nonumber\\  
m_x^2 &=& m_{1,\ldots,p-2}^2
\eea
where we have used from (3.5)
\be
f' = \sqrt{1-f^{5-p}l^2}, \quad f'' = -\frac{5-p}{2} l^2 f^{4-p}, \quad
f''' = - \frac{5-p}{2} (4-p) l^2 f^{3-p} f'
\ee
Several comments are in order here. First of all, note that the metric
is light-cone time ($x^+$) depenedent and therefore does not lead to solvable
string theories. Secondly, we comment that the scaling parameter defined by
$R^2 = \alpha' b/a^2$, with $a^{7-p} = b/(nc_pg)$, is taken to infinity
while taking the Penrose limit and in terms of the parameters of the theory
this means that it can be achieved either by (i) $b \to \infty$, $a\,=$ 
fixed, or by (ii) $a \to 0$, $b\,=$ fixed. From the relation between $a$ and
$b$ given above the condition (i) implies $g \to 0$, $n \to \infty$ with
$bg\,=$ fixed and $b^2/n\,=$ fixed. On the other hand the condition (ii)
simply implies $n \to \infty$ and $b$, $g$ fixed. For case (i) we have
large noncommutativity and for case (ii) since $a \to 0$ means $ar \to 
0$\footnote{This can also be achieved by going to the IR, where the theory
would become commutative again.}, therefore, we will not have any 
noncommutative
effect. However, we would like to emphasize that the limit for case (ii)
as mentioned above is not quite right and we will be misled if we look
at only the metric and the dilaton field. Actually, a closer look at the other
gauge fields reveal that we will not get back the commutative theory
by taking this naive limit and the proper limit along with 
$a \to 0$ and $n \to \infty$ should be $b \to 0$, $g \to \infty$ with $bg=$
fixed and $a^{(7-p)/2}/b \sim 1/\sqrt{n} \to 0$.

Let us discuss case (ii) first and then we will discuss case (i). For
case (ii) as we mentioned in section 2, the NCYM supergravity configuration
would reduce to ordinary YM theory and we also have to use the relation 
(2.10). The Penrose limit of NCYM theory also reduces to that of ordinary
D$p$-branes (in the near horizon limit). Indeed we note that since $f^{7-p}
\ll 1$ in this case, the functions $F_{p-1,p} = F_{1,\ldots,p-2} =
(f^{(7-p)/4})''/f^{(7-p)/4}$ and masses satisfy
\be
m_{p-1,p}^2 = m_{1,\ldots,p-2}^2 = \frac{7-p}{16} f^{3-p} \left[-(3-p)f^{p-5}
+ (13-3p)l^2\right]
\ee
The forms of the other masses remain the same as given in (3.14). However,
note that these formulae are not the same as given in eq.(5.19) of ref.
\cite{gps}
for ordinary D$p$-branes, because, here we have defined $ar = e^U = f(x^+)$.
The parameter `$a$' does not have any obvious meaning in the commutative 
theory. But we note that this parameter can be completely absorbed by scaling
$f \to f/a$, $l^2 \to a^{5-p}l^2$, $x^+ \to x^+/a$ and $x^- \to a x^-$. In that
case the metric (3.11) as well as the masses would take very similar form
as given in (5.18) and (5.19) of ref.\cite{gps} and the discussions of
mass$^2$ would also be the same. Unlike in that reference our formula
is valid even for the $p=5$ case.

For case (i) we have noncommutativity since $a\,=$ fixed. We will look at
the theory in the UV where $ar \gg 1$. In this region the behavior of mass$^2$
would be
\be
m_{p-1,p}^2 = \frac{7-p}{16} f^{3-p} \left[-(11-p) f^{p-5} + (p+1) l^2\right]
\ee
with all other mass$^2$'s remaining the same as given in (3.14). Also the
behavior of the dilaton is
\be
e^{\phi} \sim \bar{g}_{YM}^2 b^{\frac{3-p}{2}} f^{\frac{(7-p)(p-5)}{4}}
\ee
We note from (3.17) and (3.18) that for $p \geq 3$, all the mass$^2$'s are
always positive except $m_{p-1,p}^2$ and $e^{\phi} \ll 1$. Since for the
consistency of geodesic motion (3.5) we find,
\bea
f & \leq & \left(\frac{1}{l^2}\right)^{\frac{1}{5-p}}, \qquad {\rm for} \quad
p < 5\nonumber\\
{\rm and} \quad l^2 & \leq &1, \qquad {\rm for} \quad p = 5
\eea
we conclude that there is no region of $f$ where $m_{p-1,p}^2$ can become
positive for $p<5$. However for $p = 5$, it remains negative for all $l^2
< 1$, except for $l^2 = 1$, where $m_{4,5}^2 = 0$. We will return to the 
last case in the next section.

In case of ordinary D$p$-branes, it was shown in \cite{gps}
that when the mass$^2$ becomes 
negative, the supergravity description should be given either by the S-dual
one (for $p$ odd) or by lifting it to M-theory (for $p$ even) by an RG flow
where mass$^2$ becomes positive again. However, this does not seem to be
possible in the present case, since $e^{\phi}$ would blow up by going to the
dual frame. For $p=2$ all the mass$^2$'s become negative and so, we conclude
that the mass$^2$ becoming negative is a generic feature 
of all NCYM theories in
the Penrose limit in dimensions $3 \leq d \leq 6$ and needs a more careful 
study to understand their physical meaning. However, for $d=6$ there is
a special case where such problem does not arise and we will consider this case
in the next section.

We would like to emphasize that it would be too naive to conclude that the
appearance of scalars of negative mass$^2$ on the world-sheet would necessarily
imply the instability of the background space-time. As has been already pointed
out in \cite{gps} that these tachyons are world-sheet tachyons and they arise
in Brinkman coordinates only in the light-cone gauge (in the static gauge
the pp-wave gives rise to the world-sheet theory of massless but interacting
scalars). For time independent masses the issue of stability of the pp-wave
background has been studied in the presence of negative mass$^2$ in 
\cite{bgs,mp}. By looking at the geodesic equation of the transverse 
coordinates from the pp-wave metric in this case, it is easy to see that a test
particle moving along the geodesic will experience a repulsive tidal force
and will be pushed off to infinity with time. This means that for strings the
center of mass (or zero mode) will experience a similar force in this
background. However, this phenomenon by itself does not imply instability
if the strings move rigidly along the geodesic. But, it can be seen that
when $ m_i P_- > \frac{1}{l_s^2}$, where $m_i$ is the mass of the $i$-th 
coordinate, $P_-$ is the light-cone momentum and $l_s$ is the string length
scale, the different parts of the string experience different repulsive
tidal forces causing the string to rip apart. Morever, under the same
condition, it can be seen that truly stringy unstable modes appear, although
they are finite in number. In other words, the appearance of unstable
modes imply a string theoretic instability of the background. However, it
has been argued in \cite{mp}, that by imposing an infrared cut-off on the
coordinate $x^i$, the background can be made stable. A careful analysis
in \cite{bgs} made at the level of linearized field equations showed that the
fluctuations around the pp-wave background with negative but constant mass$^2$
are indeed stable. For the case of time dependent mass$^2$ i.e. for the case
we have discussed so far, it is in general difficult to solve the geodesic 
equations, but for the cases where they can be solved \cite{fis}, it would be 
interesting to understand the issue of stability along the lines 
\cite{bgs,mp}.  

Finally, we note that for $p=3$, the mass$^2$ relations given in (3.14) 
simplify to 
\bea
m_z^2 &=& m_1^2 = m_x^2 = l^2\nonumber\\
m_{2,3}^2 &=& \frac{l^2}{(l^4 + \sin^4(lu))^2}\left[l^8 - \sin^8(lu) - 2 
\sin^2(lu) \cos^2(lu)(\sin^4(lu) - 5l^4)\right]
\eea
where we have integrated the evolution equation (3.5) and used the form of
$f$ obtained from there as,
\be
f(u) = e^U = \frac{1}{l} \sin(lu)
\ee
By scaling $u = x^+ \to a x^+$, $x^- \to x^-/a$ and $l \to l/a$, the metric 
(3.11) reduces to 
\bea
ds^2 &=& - 4 dx^+ dx^- - l^2 \left(\vec{z}^2 + (x^1)^2 + x^2 + g(x^+) ((x^2)^2
+ (x^3)^2)\right) (dx^+)^2 \nonumber\\
& & \qquad + dx^2 + \sum_{i=1}^3 (dx^i)^2 + d\vec{z}^2
\eea
where
\be
g(x^+) = \frac{1}{(l^4 + a^4 \sin^4(lx^+))^2}\left[l^8 - a^8 \sin^8(lx^+) -
\frac{1}{2}a^4 \sin^2(2lx^+)(a^4 \sin^4(lx^+) - 5l^4)\right]
\ee
This is exactly the same form of the metric obtained in ref.\cite{hrv}. 
For $a \to 0$,
$g(x^+) \to 1$ and we get back the maximally supersymmetric pp-wave resulting
from AdS$_5 \times$S$^5$.

\sect{Penrose limit, 6d NCYM and a solvable string theory}

In this section we will discuss the Penrose limit of (D5, D3) bound state
system in the NCYM limit (this is the supergravity dual of $6d$ NCYM theory)
separately and will see that the Penrose limit in this case would lead to a
solvable string theory. The full supergravity configuration of (D3, D5)
system is given as \cite{lur,bmm},
\bea
ds^2 &=& H^{1/2}\left[H^{-1}(-(d\tilde{x}^0)^2 + \sum_{i=1}^{3}
(d\tilde{x}^i)^2) + H'^{-1}((d\tilde{x}^4)^2 + (d\tilde{x}^5)^2
) + d\tilde{r}^2 + \tilde{r}^2 d\Omega_3^2\right]\nonumber\\
e^{\phi} &=& g_s H'^{-1/2}\nonumber\\
B &=& \tan\varphi H'^{-1} d\tilde{x}^{4} \wedge d\tilde{x}^5\nonumber\\
A^{(2)} &=& n\alpha' \sin^2\theta d\psi \wedge d\phi\nonumber\\
A^{(4)} &=& -\frac{1}{2} m \alpha' H'^{-1} \sin^2\theta d\tilde{x}^4
\wedge d\tilde{x}^5 \wedge d\psi \wedge d\phi + \frac{\sin\varphi}{g_s}
H^{-1} d\tilde{x}^0 \wedge d\tilde{x}^1 \wedge d\tilde{x}^2 \wedge 
d\tilde{x}^3
\eea
where we have written $d\Omega_3^2 = \cos^2\theta d\psi^2 + d\theta^2
+ \sin^2\theta d\phi^2$. The harmonic functions $H$ and $H'$ and the angle
$\cos\varphi$ are defined in (2.2) and (2.3) with $p=5$. The NCYM limit is
given in (2.5) with $p=5$. The supergravity dual of $6d$ NCYM theory would
then be given as,
\bea
ds^2 &=& \alpha' \frac{ar}{b}\left[-(dx^0)^2 + \sum_{i=1}^{3}
(dx^i)^2 + \frac{1}{1+(ar)^2}\sum_{j=4}^5(dx^j)^2 
+ \frac{b^2}{a^2}(\frac{dr^2}
{r^2} + 
d\Omega_3^2)\right]\nonumber\\
e^{\phi} &=& g\frac{ar}{\sqrt{1+(ar)^2}}\nonumber\\
B &=& \frac{\alpha'}{b}\frac{(ar)^2}{1+(ar)^2} dx^4 \wedge dx^5\nonumber\\
A^{(2)} &=& n\alpha' \sin^2\theta d\psi \wedge d\phi\nonumber\\
A^{(4)} &=& -\frac{1}{2} n \frac{(\alpha')^2}{b}\frac{(ar)^2}{1+(ar)^2}
\sin^2\theta dx^4
\wedge dx^5 \wedge d\psi \wedge d\phi\nonumber\\
& &  + \frac{(\alpha')^2}{g b^2} (ar)^2
dx^0 \wedge dx^1 \wedge dx^2 \wedge 
dx^3
\eea
Here the fixed coordinates $x^{0,\ldots,5}$ are as defined before and the NCYM
coupling is $\bar{g}_{YM}^2 = (2\pi)^3 g b$. The parameter $a^2 = b/(ng)$,
$n$ being the number of D5-branes. The Penrose limit of this theory for
$l^2 < 1$ is discussed in section 3 and is taken by scaling the coordinates
$x^{0,\ldots,5} \to (b/a) x^{0,\ldots,5}$. We have also defined the scaling
parameter $R^2 = (\alpha' b)/a^2$, where now $b/a = \sqrt{bng} = \sqrt{
(\bar{g}_{YM}^2 n)/(2\pi)^3}$. The configuration is given in Brinkman
coordinates as,
\bea
ds^2 &=& - 4 dx^+ dx^- - \left(m_z^2 \vec{z}^2 + m_x^2 (x^2 +
\sum_{i=1}^3(x^i)^2) 
+ m_{4,5}^2 ((x^4)^2 + (x^5)^2)\right)(dx^+)^2\nonumber\\
& & \qquad\qquad\qquad\qquad\qquad\qquad\qquad\qquad + dx^2 + d\vec{z}^2 
+ \sum_{i=1}^5 (dx^i)^2
\nonumber\\
e^{\phi} &=& g\frac{f}{\sqrt{1+f^2}}\nonumber\\
B &=&  f dx^4 \wedge dx^5\nonumber\\
A^{(2)} &=& \frac{l}{gf} z^2 dx^+ \wedge d\phi\nonumber\\
dA^{(4)} &=& \frac{l}{g} dx^4 \wedge dx^5 \wedge dx^+ 
\wedge dz_1 \wedge dz_2  
\eea
where mass$^2$'s are as given in (3.14) with $p=5$. Also note that $z^2 =
\vec{z}^2 = z_1^2 + z_2^2$, where $z_1 = z \cos\phi$, $z_2 = z \sin\phi$
and therefore, $d(z^2d\phi) = 2 dz_1 \wedge dz_2$. We have defined as before
$ar = e^U = f(x^+)$, where $f$ is obtained by solving the evolution equation
(3.6) and has the form $f(x^+) = \sqrt{1-l^2} x^+$. Thus (4.3) gives
the pp-wave limit of the supergravity dual of 6$d$ NCYM theory. As we mentioned
earlier in the UV where $ar \gg 1$, even though the dilaton in (4.3) $e^\phi
\ll 1$, $m_{4,5}^2$ remains negative in this case. But all other mass$^2$'s
namely $m_x^2$, $m_z^2$ are positive. Also since the masses are light-cone
time dependent, it does not lead to solvable string theory for $l^2 < 1$. When
$ar \ll 1$, (4.3) would reduce to the Penrose limit of D5-brane (in the
near horizon limit). Note that in this case the parameter `$a$' has to
be absorbed by proper scaling as mentioned earlier in section 3.

Now looking at the evolution equation (3.5) for the general value of the
parameter $l^2$, we notice that there exists a null geodesic corresponding to
$l^2 = 1$ for which $e^U$ or $U$ remains constant. The null geodesic is now
confined  in the $(x^0,\,\psi)$ plane as in the case of maximally 
supersymmetric AdS$_5\times$S$^5$. The geodesic is $U = U_0 =$ constant,
$x^{1,\ldots,5} = \theta = 0$, $x^0 = \psi = \lambda$, where $\lambda$ is
related to the affine parameter. Now let us look at the metric (3.2) for 
$p=5$ and make the following coordinate change,
\bea
U &\to & U_0 + e^{-U_0/2} x\nonumber\\
\theta &\to & e^{-U_0/2} z\nonumber\\
x^{1,2,3} &\to & e^{-U_0/2} x^{1,2,3}\nonumber\\
x^{4,5} &\to & e^{-U_0/2} (1+ e^{2U_0})^{1/2} x^{4,5}\nonumber\\
x^0 &\to & e^{-U_0/2}(x^+ + x^-)\nonumber\\
\psi &\to & e^{-U_0/2} (x^+ - x^-)
\eea
By further rescaling the coordinates as $x^+ \to x^+$, $x^- \to x^-/R^2$,
$x \to x/R$, $z \to z/R$, $x^{1,\ldots,5} \to x^{1,\ldots,5}/R$ and taking
$R \to \infty$, the metric takes the form,
\be
ds^2 = - 4 dx^+ dx^- - \vec{z}^2 (dx^+)^2 + \sum_{i=1}^5 (dx^i)^2 + dx^2
+ d\vec{z}^2
\ee
Where in writing the above metric we have rescaled $x^{\pm} \to e^{{\pm}U_0/2}
x^{\pm}$. It is clear from (4.5) that only two of the eight bosonic 
coordinates are massive and time independent. So, this will lead to a solvable 
string theory. The masses of the bosonic fields can be made arbitrary by 
scaling $x^{\pm} \to \mu^{\pm 1} x^{\pm}$ and then the metric as well as the
other fields would take the following forms,
\bea 
ds^2 &=& - 4 dx^+ dx^- - \mu^2\vec{z}^2 (dx^+)^2 + d\vec{z}^2 + 
\sum_{i=3}^8 (dz_i)^2\nonumber\\
e^\phi &=& g \frac{e^{U_0}}{\sqrt{1+ e^{2U_0}}}\nonumber\\
B &=&  - e^{U_0} dz_3 \wedge dz_4 \qquad \Rightarrow \qquad dB\,\, =\,\, 0
\nonumber\\
F^{(3)} &=& dA^{(2)}\,\,=\,\,-\frac{2}{g} e^{-U_0} \mu dx^+ \wedge dz_1 
\wedge dz_2\nonumber\\
F^{(5)} &=& dA^{(4)} - \frac{1}{2} B \wedge F^{(3)}\nonumber\\
&=& \frac{2\mu}{g}( dx^+ \wedge dz_1 \wedge dz_2 \wedge dz_3 \wedge dz_4
+ dx^+ \wedge dz_5 \wedge dz_6 \wedge dz_7 \wedge dz_8)
\eea
Note that we have renamed the coordinates $(x^5,\,x^4,\,x^3,\,x^2,\,x^1,\,x)$
as $(z_3,\,z_4,\,z_5,\,z_6,\,z_7,\,z_8)$. It would be interesting to study the
complete quantization of the above system, but in the following we will
quantize only the bosonic closed string sector and mention a few words about
the fermionic sector for completeness. Also, we mention that for case (ii)
discussed in the previous section, 
the above configuration matches with the Penrose limit of D5-branes considered
in \cite{osa}.

The bosonic part of the Green-Schwarz action is
\be
-4\pi\alpha' S_b = \int d\tau \int_0^{2\pi\alpha'p^+} d\sigma (\eta^{ab}
G_{\mu\nu} \partial_a x^\mu \partial_b x^\nu + \epsilon^{ab} B_{\mu\nu}
\partial_a x^\mu \partial_b x^\nu)
\ee
where $\eta^{ab} = {\rm diag}(-1,\,1)$ is the world-sheet metric and
$\epsilon^{\tau\sigma} = 1$. For the background (4.6), the $B_{\mu\nu}$ term
will not contribute and so, we get,
\be
-4\pi\alpha' S_b = \int d\tau \int_0^{2\pi\alpha'p^+} d\sigma (\eta^{ab} 
\partial_a z_i \partial_b z_i + \mu^2 z_k^2)
\ee
where we have used the light-cone gauge $x^+ = \tau$ and in the above
$i=1,\ldots,8$ and $k=1,2$. The equations of motion following from (4.8)
are
\bea
\eta^{ab} \partial_a\partial_b z_l &=& 0, \qquad {\rm for} \quad l = 3,\ldots,8
\nonumber\\
\eta^{ab}\partial_a\partial_b z_k - \mu^2 z_k &=& 0, \qquad {\rm for}
\quad k=1,2
\eea
The equations of motion can be solved as usual by Fourier expanding the 
bosonic fields $z_i$, with $i=1,\ldots,8$. The bosonic part of the light-cone
Hamiltonian would be given as,
\be
2p^- = \sum_n \left(N_n^{(l)} \frac{|n|}{\alpha'p^+} + N_n^{(k)} \sqrt{\mu^2 +
\frac{n^2}{(\alpha'p^+)^2}}\right)
\ee
where $k=1,2$ corresponds to the massive bosons and $l=3,\ldots,8$ corresponds
to the massless bosons. Now in order to relate the string spectrum to the
states in NCYM theory, we write,
\be
\frac{\partial}{\partial x^+} = \frac{\partial}{\partial x^0} + \frac{\partial}
{\partial \psi}, \qquad \frac{\partial}{\partial x^-} = \frac{e^{-U_0}}{R^2}
(\frac{\partial}{\partial x^0} - \frac{\partial}{\partial \psi})
\ee
In terms of the generators of translation of original $x^0$ (before rescaling 
by $(b/a)x^0$) we find
\bea
\frac{2p^-}{\mu} &=& i \frac{\partial}{\partial x^+}\,\,=\,\, \frac{b}{a}
E - J_V\nonumber\\
2\mu p^+ &=& i \frac{\partial}{\partial x^-}\,\,=\,\, \frac{e^{-U_0}}{R^2}
(\frac{b}{a}E + J_V)
\eea
where we have used $ i \frac{\partial}{\partial x^0} = \frac{b}{a}E$ and 
$ - i \frac{\partial}{\partial \psi} = J_V$. We thus find a correspondence
of string propagation in the background (4.6) with the states in $6d$ NCYM
theory. To be precise, the string spectrum given in (4.10) corresponds to 
states in NCYM theory with energy and $U(1)$ R-charge
\be
\frac{b}{a} E, \,\,\, J_V \sim \frac{b}{a^2} \to \infty, \qquad
{\rm with} \quad \frac{b}{a} E - J_V \,\,=\,\, {\rm finite}
\ee
We note from (4.12) that for $R^2 \to \infty$,
\be
R^2 \mu p^+ = e^{-U_0} J_V
\ee
The light-cone energy therefore takes the form,
\be
\frac{2p^-}{\mu} = \sum_n \left(N_n^{(l)} e^{U_0} \frac{b}{J_V a^2} |n|
+ N_n^{(k)} \sqrt{1+ e^{2U_0}\frac{b^2}{a^4 J_V^2} n^2}\right)
\ee
We would like to mention that the NCYM supergravity configuration (4.3)
before taking the Penrose limit has $SO(4) \simeq SU(2)_L \times SU(2)_R$
isometry of S$^3$ which is the full R-symmetry group of the $6d$ NCYM theory.
The $U(1)_1 \times U(1)_2 \equiv U(1)_V \times U(1)_A$ subgroup of 
this full isometry group corresponds
to the isometry of $\psi$ and $\phi$ in $d\Omega_3^2$. If we write the two 
bosonic fields $z_1$ and $z_2$ in terms of complex scalars $z = 
(z_1 + i z_2)/2$ and $\bar{z} = (z_1 - i z_2)/2$, then they will carry $U(1)_2
= U(1)_A$ charge corresponding to the angular coordinate $\phi$. On 
the other hand
the massless scalars $z_3,\ldots,z_8$ 
are $U(1)_A$-charge neutral, where $U(1)_{V(A)}$ denotes the vector(axial
vector) subgroup of $U(1)_L \times U(1)_R$. We thus find that the states with
non-zero $U(1)_A$ charge have different energies than those with vanishing
$U(1)_A$-charge. A similar conclusion has been drawn in \cite{osa} for 
the case of $6d$
``gauge'' theory, the holographic dual of (NS5, D5) bound state (in the
OD5 limit), although the details are quite different.

The fermionic part of the Green-Schwarz action comes from the direct
covariantization of the quadratic fermionic terms of the flat space action
\cite{gres} and has the form \cite{rusey}
\be
-4\pi\alpha'S_f = i \int d\tau \int_0^{2\pi\alpha'p^+} d\sigma \big(\eta^{ab}
\delta_{IJ} - \epsilon^{ab} \rho_{3IJ}\big)\partial_a x^\mu \theta^I \Gamma_\mu
({\cal D}_b \theta)^J
\ee
where the supercovariant derivative is defined as,
\bea
{\cal D}_b &=& \partial_b + \frac{1}{4} \partial_b x^\nu 
\left[\big(\omega_{\hat{\lambda}
\hat{\rho}\nu} - \frac{1}{2} H_{\hat{\lambda}\hat{\rho}\nu} \rho_3\big) 
\Gamma^{\hat{\lambda}\hat{\rho}}\right.
\nonumber\\
& & \qquad\qquad\qquad\qquad \left.- e^\phi \big(\frac{1}{3!}
F_{\hat{\lambda}\hat{\rho}\hat{\sigma}}\Gamma^{
\hat{\lambda}\hat{\rho}\hat{\sigma}} \rho_1 + \frac{1}{2(5!)}F_{\hat{\mu}_1
\ldots\hat{\mu}_5} 
\Gamma^{\hat{\mu}_1\ldots\hat{\mu}_5}\rho_0\big)\Gamma_\nu\right] 
\eea
In the above $I,J = 1,2$ are the labels of two real MW spinors. 
The hatted indices are the tangent space indices. $\rho$'s in 
$I,J$ space are related to the Pauli matrices as $\rho_3 = {\rm diag}\,\,
(1,\,-1)$, $\rho_1 = \sigma_1$ and $\rho_0 = i \sigma_2$. 
$\Gamma^{\hat{\mu}}$'s
are 32 $\times$ 32 Dirac matrices satisfying the Clifford algebra
\be
\{\Gamma^{\hat{\mu}},\, \Gamma^{\hat{\nu}}\} = 2 \eta^{\hat{\mu}\hat{\nu}}
\ee
where $\eta^{\hat{\mu}\hat{\nu}}$ is the mostly positive Lorentzian metric and
$\Gamma^{\hat{\mu}_1\ldots}$ is as usual the totally antisymmetric product of
gamma matrices. In the light-cone gauge we set $x^+ = \tau$ and $\Gamma^+
\theta^I = 0$\footnote{After rescaling $x^- \to -(1/2) x^-$ in the background
(4.6) it is easy to verify that for the upper $+$ and lower $-$ indices, it
is not necessary to distinguish between the hatted and the unhatted indices.
The same is true for the transverse indices $i = 1, \ldots,8$. So, in the 
following all the indices are hatted except the lower $+$ and upper $-$ 
indices and we will remove the hats while writing the indices explicitly.}. 
In this gauge the 
non-zero contribution to the action comes
only when both the `internal' and the `external' $\partial_a x^\mu$ factor
of the action become $\delta_+^\mu \delta_a^0$. For the background (4.6)
(with a rescaling $x^- \to -(1/2) x^-$) we can calculate
\be
\omega_{\hat{\lambda}\hat{\rho}\mu}\Gamma^{\hat{\lambda}\hat{\rho}}
= - 2 \mu^2 (z_1 \Gamma^{+1} + z_2 \Gamma^{+2})
\ee
and so the supercovariant derivative simplifies to 
\bea
{\cal D}_0 &=& D_0 - \frac{1}{4} e^{\phi}\left(F_{+12} \Gamma^{+12}
\rho_1 + \frac{1}{2}F_{+1234}
(\Gamma^{+1234} + \Gamma^{+5678})\rho_0\right) 
\Gamma_+\nonumber\\
{\cal D}_1 &=& \partial_1
\eea
where $D_0 = \partial_0 - \frac{1}{2}\mu^2(z_1 \Gamma^{+1} + z_2 \Gamma^{+2})$.
So, the action would take the form,
\bea
-4\pi\alpha' S_f &=& i\int d\tau \int_0^{2\pi\alpha'p^+} d\sigma
\left[-\delta_{IJ}
\theta^I \Gamma_+\left(D_0 - \frac{e^{\phi}}{4}(F_{+12}\Gamma^{+12}\rho_1
\right.\right.\nonumber\\
& &\left.\left.+ \frac{1}{2}F_{+1234}(\Gamma^{+1234} + 
\Gamma^{+5678})\rho_0)\Gamma_+\right)^{JK}\theta_K
-\rho_{3IJ} \theta^I \Gamma_+ 
\partial_1^{JK} \theta_K\right]
\eea
One can write down the 
equations of 
motion from the above action and quantize the system in a straightforward
way. We will not study the quantization of the fermionic sector in any
further detail here. However, we will show that pp-wave background of the 
$6d$ NCYM theory given in (4.6) preserves only half of the space-time 
supersymmetry.

Now in order to study the supersymmetry of the above background we follow
closely the analysis made in ref.\cite{bfhp}. We first write the 
supersymmetry variations of the dilatino and the gravitino as follows
\cite{sch},
\bea
\delta\chi &=& \left[\Gamma^\mu \partial_\mu \phi - \frac{1}{4(3!)}e^{\phi}
F_{\hat{\lambda}\hat{\rho}\hat{\nu}}\Gamma^{\hat{\lambda}\hat{\rho}\hat{\nu}}
\rho_1\right]\epsilon\\
\delta \psi_\mu &=& {\cal D}_\mu \epsilon \equiv
\left[\partial_\mu + \frac{1}{4}  
\big(\omega_{\hat{\lambda}
\hat{\rho}\mu} - \frac{1}{2} H_{\hat{\lambda}\hat{\rho}\mu} \rho_3\big) 
\Gamma^{\hat{\lambda}\hat{\rho}}\right.
\nonumber\\
& & \qquad\qquad\qquad\qquad \left.- \frac{e^\phi}{4} \big(\frac{1}{3!}
F_{\hat{\lambda}\hat{\rho}\hat{\sigma}}\Gamma^{
\hat{\lambda}\hat{\rho}\hat{\sigma}} \rho_1 + \frac{1}{2(5!)}F_{\hat{\mu}_1
\ldots\hat{\mu}_5} 
\Gamma^{\hat{\mu}_1\ldots\hat{\mu}_5}\rho_0\big)\Gamma_\mu\right]\epsilon
\eea 
where $\epsilon = \left(\begin{array}{c}\epsilon_1\\ \epsilon_2\end{array}
\right)$ is a two component spinor on which the matrices $\rho_0$ and 
$\rho_1$ act and satisfy the chirality condition $\Gamma_{11} 
\epsilon_{1,2} \equiv \Gamma^{+-}\Gamma^1 \ldots \Gamma^8 \epsilon_{1,2}
= - \epsilon_{1,2}$. Since the background is bosonic, for consistency we
set $\delta\chi = \delta\psi_\mu = 0$. So, from (4.22) we get,
\be
\left[\Gamma^\mu\partial_\mu \phi - \Gamma^{+12} \Lambda\right]\epsilon
= 0
\ee
where we have defined $\Lambda = (e^\phi/4) F_{+12}\rho_1 = \lambda \rho_1$.
Note that for the background (4.6) the dilaton is constant and therefore
the above equation can be satisfied if $\Gamma^{+12} \epsilon = 0$.
Furthermore, from (4.23) we obtain,
\be
\left[\partial_\mu + \frac{1}{4}  
\omega_{\hat{\lambda}
\hat{\rho}\mu}  
\Gamma^{\hat{\lambda}\hat{\rho}}
- \frac{e^\phi}{4} \big(\frac{1}{3!}
F_{\hat{\lambda}\hat{\rho}\hat{\sigma}}\Gamma^{
\hat{\lambda}\hat{\rho}\hat{\sigma}} \rho_1 + \frac{1}{2(5!)}F_{\hat{\mu}_1
\ldots\hat{\mu}_5} 
\Gamma^{\hat{\mu}_1\ldots\hat{\mu}_5}\rho_0\big)\Gamma_\mu\right]\epsilon=0
\ee
In components it gives $\partial_-\epsilon = 0$ and so, $\epsilon$ is 
independent of $x^-$ coordinate. For the $i$-th component $(i=1,\ldots,8)$,
we get
\be
(\partial_i + \Omega_i)\epsilon = 0
\ee
where,
\be
\Omega_i = \cases{\Gamma^+\Gamma^i(I\Lambda + J\Lambda'), & {\rm for}
$\quad i=1,2$\cr
\Gamma^+ \Gamma^i(-I\Lambda + J\Lambda'), & {\rm for} $\quad i=3,4$\cr
\Gamma^+ \Gamma^i (-I\Lambda + K\Lambda'), & {\rm for} $\quad 
i= 5,\ldots,8$\cr}
\ee
Here $I=\Gamma^{12}$, $J=\Gamma^{1234}$, $K=\Gamma^{5678}$, $\Lambda'
= (e^\phi/4)F_{+1234}\rho_0 = \lambda'\rho_0$. We also note the identities,
$I^2=-1$, $J^2=K^2=1$, $IJ=JI$, $IK=KI$, $JK=KJ$ and $\Lambda^2 = \lambda^2$,
$\Lambda'^2 = - \lambda'^2$, $\Lambda\Lambda'-\Lambda'\Lambda=-2\lambda\lambda'
\rho_3$. Now since, $\Omega_i\Omega_j=0$ for all $i,j$, so we get from (4.26)
\be
\epsilon = (1-\sum_{i=1}^8 z_i \Omega_i)\psi
\ee
where the spinor $\psi$ depends only on $x^+$ coordinate. Finally, the
$+$ component of the killing spinor equation (4.25) gives,
\be
\left[\partial_+ - \frac{\mu^2}{2}\sum_{i=1}^2 z_i \Gamma^+ \Gamma^i -
(2I\Lambda + J\Lambda' + K\Lambda')\right]\epsilon = 0
\ee
Substituting (4.28) in (4.29) and using the identities mentioned above 
eq.(4.29) can be simplified in the following form
\bea
&& \partial_+ \psi - (2I\Lambda +(J+K)\Lambda')\psi\nonumber\\
&& = \sum_{i=1}^2 z_i (\frac{\mu^2}{2} - 4 \lambda^2 - 2 \lambda'^2 +
2IK\lambda\lambda'\rho_3)\Gamma^i\Gamma^+ \psi\nonumber\\
&& + \sum_{i=3}^4 2z_i (I(K-2J)\lambda\lambda'\rho_3 -  \lambda'^2) 
\Gamma^i\Gamma^+ \psi\nonumber\\
&& + \sum_{i=5}^8 2z_i (I(J-2K)\lambda\lambda'\rho_3 -  \lambda'^2) 
\Gamma^i\Gamma^+ \psi
\eea
Since the lhs of (4.30) is independent of $z_i$ coordinates whereas the rhs
depends on $z_i$, they must be separately zero. It is clear that the rhs
of (4.30) can vanish only if $\Gamma^+ \psi = 0$ i.e. only half of the
components of $\psi$ are non-trivial indicating that the background
is half supersymmetric. This is also consistent with dilatino equation
(4.24). The exact form of the killing spinor $\epsilon$ can be obtained by 
putting the lhs of (4.30) to zero, solving this for $\psi$ and substituting 
it in (4.28). This therefore
concludes our discussion of Penrose limit of $6d$ NCYM theory.

\sect{Conclusion}

In this paper we have studied the Penrose limit of the supergravity duals
of NCYM theories in dimensions $3\leq d \leq 6$. The supergravity 
descriptions are obtained from (D$(p-2)$, D$p$) bound state configurations
for $2\leq p \leq 5$ of type II string theories in the so-called NCYM limit.
We found that the Penrose limit in these cases gives a one parameter ($l$)
family of string theories in a time dependent pp-wave background. Because
of the time dependence the corresponding string theories are not exactly
solvable. We also obtained the expressions of mass$^2$'s of various bosonic
fields. We found that for $p\geq 3$, all the mass$^2$'s remain positive
except the ones associated with the bosonic fields corresponding to the 
noncommutative coordinates. The mass$^2$'s for the latter coordinates are
always negative. However, for $p=2$, not only the mass$^2$'s of the bosonic 
fields of the noncommutative directions, but also those associated with the
commuting directions are negative. So, we conclude that mass$^2$ becoming 
negative is a generic feature of the NCYM theories in the various dimensions
we studied and requires a careful study to understand their physical meaning.
In particular, it has been pointed out in \cite{gps}, that for such 
time dependent 
pp-wave background there is a map bewteen the associated time dependent 
quantum mechanical problem and the RG flow in the dual gauge theory. It should
be interesting to understand the precise meaning of this map in the present
context.

In obtaining the Penrose limit we have defined a scaling parameter 
$R^2 = \alpha'b/a^2 \to \infty$, in terms of the known parameters of the 
NCYM theories. $R^2 \to \infty$ means either (i) $b \to \infty$, $a=$ fixed,
or (ii) $a \to 0$, $b \to 0$, $n \to \infty$, $g \to \infty$ with $bg$
fixed and $a^{(7-p)/2}/b \sim 1/\sqrt{n} \to 0$, where $b$ is the 
noncommutativity parameter,
$n$ is the number of D$p$-branes and $bg \sim
\bar{g}_{YM}^2$, is the gauge coupling. For case (i) there is a large
noncommutativity and all our discussions above apply to this case. For case
(ii) the theories become commutative. We have shown how for case (i)
our results reproduce the Penrose limit of four dimensional NCYM theory
studied in \cite{hrv}. We have also shown that for case (ii) 
we recover the Penrose 
limit of ordinary D$p$-branes (in the YM limit) studied in \cite{gps}.

All the discussions above apply to the NCYM theories in $3 \leq d \leq 6$
for the generic value of the parameter $l$. We found that in $d=6$, there
exists another Penrose limit of NCYM theory, when the parameter $l$
is saturated i.e. $l=1$, which leads to a solvable string theory. We have
studied the quantization of this string theory in the complete NSNS and
RR pp-wave background. We studied the quantization of the bosonic sector in
detail and pointed out some features of the fermionic sector where
we have also shown that the pp-wave background is half supersymmetric. 
We have obtained
the spectrum of the light-cone Hamiltonian of the string theory and
mentioned their relations to the states in $6d$ NCYM theory.

\section*{Acknowledgements}

We would like to thank Alok Kumar for discussions and collaboration at an
early stage of this work.


\begin{thebibliography}{99}

\bibitem{bfhp} M. Blau, J. Figueroa-O'Farrill, C. Hull and G. Papadopoulos,
``A new maximally supersymmetric background of IIB superstring theory'',
JHEP 01 (2002) 047, [hep-th/0110242].

\bibitem{met} R. Metsaev, ``Type IIB Green-Schwarz superstring in plane wave
Ramond-Ramond background'', Nucl. Phys. B625 (2002) 70, [hep-th/0112044].

\bibitem{metsey} R. Metsaev and A. Tseytlin, ``Exactly solvable model of 
superstring in plane wave Ramond-Ramond background'', hep-th/0202109.

\bibitem{rusey} J. Russo and A. Tseytlin, ``On solvable models of type IIB
superstring in NS-NS and R-R plane wave backgrounds'', JHEP 04 (2002) 021,
[hep-th/0202179].

\bibitem{pen} R. Penrose, ``Any space-time has plane wave as a limit'',
in Differential Geometry and Relativity, Reidel, Dordrecht, 1976.

\bibitem{guv} R. G\"uven, ``Plane wave limits and T-duality'', Phys. Lett.
B482 (2000) 255, [hep-th/0005061].

\bibitem{blfhp} M. Blau, J. Figueroa-O'Farrill, C. Hull and G. Papadopoulos,
``Penrose limits and maximal supersymmetry'', hep-th/0201081.

\bibitem{mal} J. Maldacena, ``The large N limit of superconformal field
theories and supergravity'', Adv. Theor. Math. Phys. 2 (1998) 231,
[hep-th/9711200].

\bibitem{agmoo} O. Aharony, S. Gubser, J. Maldacena, H. Ooguri and
Y. Oz, ``Large N field theories, string theory and gravity'', Phys.
Rep. 323 (2000) 183, [hep-th/9905111].

\bibitem{gkp} S. Gubser, I. Klebanov and A. Polyakov, ``Gauge theory 
correlators from non-critical string theory'', Phys. Lett. B428 (1998)
105, [hep-th/9802109].

\bibitem{witt} E. Witten, ``Anti de Sitter space and holography'', Adv.
Theor. Math. Phys. 2 (1998) 253, [hep-th/9802150].

\bibitem{bmn} D. Berenstein, J. Maldacena and H. Nastase, ``Strings in flat
space and pp waves from N = 4 super Yang-Mills'', JHEP 04 (2002) 013, 
[hep-th/0202021].

\bibitem{imsy} N. Itzhaki, J. Maldacena, J. Sonnenschein and S. Yankielowicz,
``Supergravity and large N limit of theories with sixteen supercharges'',
Phys. Rev. D58 (1998) 046004, [hep-th/9802042].

\bibitem{bfp} M. Blau, J. Figueroa-O'Farrill and G. Papadopoulos, ``Penrose
limits, supergravity and brane dynamics'', hep-th/0202111.

\bibitem{ps} L. Pando Zayas and J. Sonnenschein, ``On Penrose limits and
gauge theories'', JHEP 05 (2002) 010, [hep-th/0202186].

\bibitem{gps} E. Gimon, L. Pando Zayas and J. Sonnenschein, ``Penrose limits
and RG flows'', hep-th/0206033.

\bibitem{chkw} R. Corrado, N. Halmagyi, K. Kennaway and N. Warner, ``Penrose
limits of RG fixed points and PP-waves with background fluxes'', 
hep-th/0205314.

\bibitem{bjlm} D. Brecher, C. Johnson, K. Lovis and R. Myers, ``Penrose
limits, deformed pp-waves and the string duals of N=1 large $n$ gauge theory'',
hep-th/0206045.

\bibitem{fis} H. Fuji, K. Ito and Y. Sekino, ``Penrose limit and string
theories on various brane backgrounds'', hep-th/0209004.

\bibitem{nw} C. Nappi and E. Witten, ``A WZW model based on non semi-simple
group'', Phys. Rev. Lett. 71 (1993) 3751, hep-th/9310112].

\bibitem{go} J. Gomis and H. Ooguri, ``Penrose limit of N = 1 gauge theories'',
Nucl. Phys. B635 (2002) 106, [hep-th/0202157].

\bibitem{kirp} E. Kiritsis and B. Pioline, ``Strings in homogeneous 
gravitational waves and null holography'', hep-th/0204004.

\bibitem{hrv} V. Hubeny, M. Rangamani and E. Verlinde, ``Penrose limits and
non-local theories'', hep-th/0205258.

\bibitem{hashi} A. Hashimoto and N. Itzhaki, ``Non-commutative Yang-Mills
and the AdS/CFT correspondence'', Phys. Lett. B465 (1999) 142, 
[hep-th/9907166].

\bibitem{mr} J. Maldacena and J. Russo, ``Large N limit of noncommutative
gauge theories'', JHEP 09 (1999) 025, [hep-th/9908134].

\bibitem{gmss} R. Gopakumar, S. Minwalla, N. Seiberg and A. Strominger,
``(OM) theory in diverse dimensions'', JHEP 08 (2000) 008, [hep-th/0006062].

\bibitem{luroy} J. X. Lu and S. Roy, ``An SL(2,Z) multiplet of type IIB 
super five-branes'', Phys. Lett. B428 (1998) 289, [hep-th/9802080].

\bibitem{aor} M. Alishahiha, Y. Oz and J. Russo, ``Supergravity 
and light-like noncommutativity'', JHEP 09 (2000) 002, [hep-th/0007215].

\bibitem{mitro} I. Mitra and S. Roy, ``(NS5, D$p$) and (NS5, D$(p+2)$, D$p$)
bound states of type IIB and type IIA string theories'', JHEP 02 (2001) 026,
[hep-th/0011236].

\bibitem{osa} Y. Oz and T. Sakai, ``Penrose limit and six dimensional 
gauge theories'', hep-th/0207223.

\bibitem{aliku} M. Alishahiha and A. Kumar, ``PP-waves from non-local 
theories'', hep-th/0207257.

\bibitem{aos} M. Alishahiha, Y. Oz and M. Sheikh-Jabbari, ``Supergravity and
large N noncommutative field theories'', JHEP 11 (1999) 007, [hep-th/9909215].

\bibitem{lur} J. X. Lu and S. Roy, ``(p+1) dimensional noncommutative 
Yang-Mills and D(p-2) branes'', Nucl. Phys. B579 (2000) 229.

\bibitem{bmm} J. Breckenridge, G. Michaud and R. Myers, ``More D-brane
bound states'', Phys. Rev. D55 (1997) 6438, [hep-th/9611174].

\bibitem{cp} M. Costa and G. Papadopoulos, ``Superstring dualities and 
$p$-brane bound states'', Nucl. Phys. B510 (1998) 217, [hep-th/9612204].

\bibitem{bgs} D. Brecher, J. Gregory and P. Saffin, ``String theory and 
classical stability of plane-waves'', hep-th/0210308.

\bibitem{mp} D. Marolf and L. Pando-Zayas, ``On the singularity structure
and stability of plane-waves,'' JHEP 01 (2003) 076, [hep-th/0210309].

\bibitem{gres} M. Green and J. Schwarz, ``Covariant description of 
superstrings'', Phys. Lett. B136 (1984) 367.

\bibitem{sch} J. Schwarz, ``Covariant field equations of chiral N=2, D=10
supergravity'', Nucl. Phys. B226 (1983) 269. 
\end{thebibliography}
\end{document}